\documentstyle[revtex-2]{aps-2}
\begin{document}
\draft
\def\half{\frac{1}{2}}
\def\psibar{\overline{\psi}}
\def\Ndangle{d\Omega_{N-2}}
\def\Ameasure{{d A_{N-2}\over{(2\pi)^{N-2}}}}
\def\dX2{d\vec{x}\cdot d\vec{x}}
\def\di{\partial}
\def\be{\begin{equation}}
\def\ee{\end{equation}}
\def\del{\nabla}
\def\da{\Delta}
\newcount\sectionnumber
\sectionnumber=0
\begin{title}
Infinite Blueshift of Charged Null Particles
\end{title}
\author{R.B. Mann$^{(1,2)}$ and W.N. Sajko$^{(1)}$}
\begin{instit}
(1) Department of Physics, University of Waterloo, Waterloo, Ontario, Canada,
N2L 3G1
\end{instit}
\begin{instit}
(2) Department of Applied Mathematics and Theoretical Physics,
Cambridge University, Cambridge, U.K. CB2 9EW
\end{instit}
\begin{center}\today\end{center}
\begin{center}{WATPHYS TH-94/05}\end{center}
\begin{abstract}
We demonstrate that charged null particles can be infinitely
blue\-shifted
in a Kerr-Newman spacetime. The surface of infinite blueshift can
be outside of the ergosphere in a Kerr-Newman spacetime, and outside
of the outer event horizon for a Reissner-Nordstrom spacetime.
Implications for extensions of the standard model which incorporate
charged neutrinos are discussed.
\end{abstract}

\widetext

The study of massless charged particles has been of increasing
interest in recent years for a number of reasons. The minimal
standard SU(3)$\times$SU(2)$\times$U(1) model (MSM)
successfully describes all
known non-gravitational interactions between quarks and leptons,
taking all three generations of neutrinos to be massless and neutral
under the U(1) of electromagnetism. However the MSM does not require
that all neutrinos be electromagnetically neutral \cite{Volkas},
and a number of authors have
begun to explore the phenomenological consequences of neutrinos that
have non-vanishing electric charges \cite{Volkas,neutchg}. It has very
recently been suggested that such charged null particles could provide
a novel resolution to the solar neutrino problem \cite{solar}.

The evolution of an electrically charged null fluids has also been a subject of
interest in the context of gravitational collapse\cite{vaidya,nulflu}.
In the spherically
symmetric case there exists an exact solution to the Einstein-Maxwell
equations\cite{vaidya} which describes such a fluid
\be
ds^{2} = 2 dr dv -\left(1-\frac{2M(v)}{r}+\frac{Q^{2}(v)}{r^{2}}\right)dv^2
+r^{2}\left(d{\theta}^{2}+\sin^{2}{\theta}d{\phi}^{2}\right) \label{e1}
\ee
which is a generalization  of the Reissner-Nordstrom and Vaidya solutions.
Here $M(v)$ and $Q(v)$ correspond respectively to the total mass and
electric charge enclosed within a shell labelled by the comoving, ingoing
null coordinate $v$.

Such a fluid may be regarded as a stream of massless charged particles
which interact only with the gravitational and electromagnetic fields.
Ori has pointed out\cite{ori} that unless such particles obey the Lorentz
force equation
\be
        k^{\mu}\del_{\mu}k^{\alpha} = q F^{\alpha}_{\ \beta}k^{\beta},
        \label{e2}
\ee
then they can penetrate into regions where the weak energy condition is
violated\cite{kamin}.  Here $k^{\alpha}$ is the null
four-momentum of a particle of charge ${q}$, and $F^{\alpha}_{\ \beta}$
is the Faraday tensor.  This equation governs the
charged null particle's trajectory and electromagnetic interactions
in spacetime in a manner analogous to that of the more well-known equations
of motion of a massive charged test particle. Equation (\ref{e2}) has been
shown to be consistent\cite{ori} with the Einstein-Maxwell field equations of a
charged null fluid with stress energy $T^{\mu\nu} = \rho k^\mu k^\nu$.

We point out in this paper that the equations of motion (\ref{e2})
necessarily imply that outgoing charged massless particles will, under certain
conditions, undergo an infinite blueshift as seen by static
observers in a Kerr-Newman spacetime. This phenomenon will occur for any
null particle of a given charge $q$ (opposite in sign to that of the
Kerr-Newman spacetime) at sufficiently low energies. This
general-relativistic infrared effect provides a qualitatively new theoretical
problem (on par with the cancellation of infrared divergences \cite{infrared})
that all models of massless charged particles must address.

We use coordinates in which the Kerr-Newman metric is
$$
ds^{2}=-(\frac{\da-a^{2}\sin^{2}\theta}{\Sigma})dt^{2}
-\frac{2a\sin^{2}\theta}{\Sigma}(r^{2}+a^{2}-\da)dtd\phi
$$
\be
+(\frac{(r^{2}+a^{2})^{2}
-a^{2}\sin^{2}\theta\da}{\Sigma})\sin^{2}\theta d\phi^{2}
+\frac{\Sigma}{\da}dr^{2}+\Sigma d\theta^{2}    \label{e3}
\ee
where
\be
\Sigma=r^{2}+a^{2}\cos^{2}\theta ,\ \ \ \ \da=r^{2}+a^{2}+Q^{2}-2Mr
\label{e4}
\ee
and where $a=\frac{J}{M}$ is the angular momentum per unit mass of the source.
The electromagnetic vector potential is
\be
A_{\mu}= -\frac{Qr}{\Sigma}(1,0,0,-a\sin^{2}\theta) \label{e5}
\ee
where $Q$ is the charge of the source.

Consider an outgoing charged null particle whose equations of motion
are given by (\ref{e2}), where $k^\mu=(k^t,k^r,k^\theta,k^\phi)$. It
is straightforward (although somewhat tedious) to check that the
general solution to (\ref{e2}) is given by
\begin{eqnarray}
        \Sigma\da k^{\phi}\sin^2(\theta)&=& a \sin^2(\theta)F(r)
          +\da(L-aE\sin^2(\theta) ) \label{e9a}\\
        \Sigma\da k^{t}&=&(r^{2}+a^{2})F(r)+a\da(L-aE \sin^2(\theta))
\label{e9b} \\
\Sigma k^{r}&=& \pm
   \sqrt{F^{2}(r)-\da[(L-aE)^{2}+{\cal Q}]}  \label{e9c}\\
\Sigma k^{\theta}&=& \pm
   \sqrt{{\cal Q}+\cos^2(\theta)[a^2
E^{2}-\frac{L^2}{\sin^2(\theta)}]}
\label{e9d}
\end{eqnarray}
where the null condition $g(k,k)=0$ has been applied, and
where
\be
        F(r)=E(r^{2}+a^{2})-aL- {q}Q r  \qquad . \label{e10}
\ee
Here ${\cal Q}$ is a constant of the motion corresponding to Carter's
fourth invariant, and $E$ and $L$ are constants of the motion defined
by
\be
g(\xi,k+{q}A)=-E, \ \ \ g(\zeta,k+{q}A)=L, \label{e8}
\ee
since $\xi=\partial/\partial t$ and
$\zeta=\partial/\partial \phi$ are Killing vectors. The plus sign in
(\ref{e9c}) corresponds to an outgoing null particle.

$E$ and $L$ respectively
correspond to the energy and z-component of angular momentum of the
charged null particle
as measured by a distant static observer with 4-velocity
$u^\mu = \xi^\mu/\sqrt{g(\xi,\xi)}$. It is straightforward to show
that a static observer located at radial coordinate $r$ with
4-velocity $u(r)$ will, as $r\to\infty$, measure charged null
particles to
have (in units of $\hbar$) a frequency
$\omega = E = -\lim_{r\to\infty}g(u(r),k)$ that is equivalent to uncharged
null particles with the same $E$. For finite values of $r$ there is
(relative to the large-$r$ limit) a frequency shift downward (upward)
for $qQ>0$ ($qQ<0$).

Consider a charged null particle
emitted at coordinate $r=r_e$ with frequency $\omega = g(u(r_e),k)
\equiv \omega_e$. For simplicity we consider motion at constant
$\theta = \theta_0$.  An observer located at $r=r_0$ will observe the
particle to have the frequency $g(u(r_0),k)\equiv \omega_o$, where from
eqs. (\ref{e9a})--(\ref{e9d}) we find
\be
\omega_e
= \frac{E-\frac{Qqr_e}{\Sigma(r_e)}}{E-\frac{Qq r_o}{\Sigma(r_o)}}
\sqrt{\frac{g_{tt}(r_o)}{g_{tt}(r_e)}}\quad \omega_o
\ee
If the emission location $r_e$ is chosen so that $g_{tt}(r_e) = 0$
({\it ie}. at the surface of the ergosphere) then
the charged null particles are infinitely redshifted just as their neutral
counterparts would be: arbitrarily large frequencies emitted near an
ergosphere will be observed to have finite values at large $r$.
However if the emission location is chosen so that
$E-\frac{Qq r_e}{\Sigma(r_e)} = 0$, or
\be
r_e = r_b \equiv \frac{qQ + \sqrt{q^2 Q^2 - 4 E^2 a^2 \cos^2(\theta_0)}}{2E}
\label{e12}
\ee
then the particles are infinitely {\it blueshifted}: arbitrarily small
frequencies emitted near $r_e$ are observed to have finite
values at large $r$.

Note that there is a surface of infinite blueshift only for $qQ >0$.
Physically this corresponds to electromagnetic repulsion of the
outgoing charged null particle by the source.  If $a\neq 0$
then the blueshift effect takes place only outside of a conical region
of angle $\hat{\theta}$ about the north and south poles where
\be
|\cos(\hat{\theta})| = \frac{qQ}{2Ea}  \qquad .  \label{e13}
\ee
The surface of infinite blueshift will occur outside of the ergosphere
whenever
\be
\frac{q}{E} \ge
\frac{M}{Q} \left(\frac{1 + 2\frac{a^2}{Q^2}\cos^2(\hat{\theta}) +
\sqrt{1 - \frac{Q^2}{M^2} - \frac{a^2}{M^2}\cos^2(\hat{\theta})}}
{1 + \frac{a^2}{Q^2}\cos^2(\hat{\theta})}\right) \qquad .\label{e14}
\ee
For a non-extremal Reissner Nordstrom spacetime this reduces to the
condition
\be
\frac{q}{E} \ge \frac{M}{Q}\left(1 + \sqrt{1 - \frac{Q^2}{M^2}}\right)
\ge 1 \label{e15}
\ee
for the surface of infinite blueshift to be outside of the outer
event horizon.

These two effects will cancel out whenever $g(\xi,\xi)$
and $g(\xi,qA)+E$ simultaneously vanish.  In this case a charged null
particle emitted at finite frequency near an ergosphere will be
observed with finite frequency at large $r$.
For the Reissner Nordstrom case this takes place for $q=E$ and $Q=M$.

Observers at large $r_o \rightarrow \infty$ will
always observe a finite frequency, denoted here by $E$.  For charged
(or neutral) null particles emitted at finite frequency near a surface
of  infinite redshift, $E$ can be arbitrarily small.  For charged
null particles emitted at finite frequency near a surface of infinite
blueshift, $E$ is bounded from below by $|g(\xi,qA(r_e))|$.

Note that the blueshift effect occurs in the infrared regime:
for sufficiently
small $E$ the charged null particle will always have a surface of
infinite blueshift outside of the event horizon. For the models
discussed in refs. \cite{Volkas,neutchg} (\ref{e15}) yields the
relation
\be
      E < \epsilon 10^{18} GeV   \label{e16}
\ee
where $q=\epsilon |e|$, where $e$ is the charge of the electron. For
muon neutrinos, $\epsilon$ can be as large as $10^{-6}$ \cite{Volkas},
yielding an extremely large threshold energy below which infinite
blueshifts can take place.
For electron neutrinos the bounds are more stringent ($\epsilon <
10^{-21}$ \cite{Marin}); however even this limit yields only
$E < 1 MeV$.

For zero-angular-momentum observers (whose 4-velocities are
$u^\mu = \nabla^\mu(t)/\sqrt{g(\nabla t,\nabla t)}$) we find that the
the frequency ratio now depends
upon the angular momentum $L$. However for any given $L$, there is
always a region of infinite blueshift outside of the ergosphere for
sufficiently small $E$.

To summarize, the Einstein-Maxwell equations require that massless charged
particles obey the equation of motion (\ref{e2}) with a Lorentz force
term\cite{ori}. Such a term accounts for electromagnetic attraction/repulsion
between charged null particles and whatever electromagnetic fields may exist
in spacetime. For the charged Vaidya solution this effect guarantees that a
charged null fluid satisfies the weak energy condition\cite{ori}. However
the same effect yields a surface of infinite blueshift about any
Kerr-Newman black hole (this effect also occurs for geometric
forces\cite{Saj} in $(1+1)$-dimensions, in which matter is minimally
coupled to a gauge field whose field strength is proportional to the area
two-form\cite{Cang}). The implications of such an effect for both
the phenomenology of massless charged particles and for the
stability of a spacetime containing them remain
interesting subjects for future work.
$$
$$

This work was supported by the Natural Sciences and Engineering Research
Council of Canada. We would like to thank K. Lake and D. Garfinkle for
interesting discussions. One of us (R.B.M.) would like to thank
S. Morsink and D. Page for correspondence which led to a revised
version of this paper.

\end{document}